\PassOptionsToPackage{hyphens}{url}
\RequirePackage[T1]{fontenc}
\documentclass[9pt,conference]{IEEEtran}

\usepackage{amsmath,amssymb}
\usepackage{graphicx}
\usepackage{booktabs}
\usepackage{multirow}
\usepackage{array}
\usepackage{tabularx}
\usepackage{makecell}
\usepackage{siunitx}
\usepackage[table]{xcolor}
\usepackage{cite}
\usepackage{tikz}
\usetikzlibrary{positioning,arrows.meta}
\usepackage{algorithm}
\usepackage[noend]{algpseudocode}

\definecolor{plUpOn}{HTML}{881F5B}     %
\definecolor{plUpBeside}{HTML}{C88EAD} %
\definecolor{plHbmOn}{HTML}{285F91}    %
\definecolor{plHbmBeside}{HTML}{5D8BB6}%

\usepackage[hidelinks]{hyperref}

\newcommand{\chen}[1]{#1}
\newcommand{\SysName}{MPU}
\hypersetup{pdftitle={Upright DRAM on Logic: Edge-Connected Memory Cubes for Bandwidth-Bound LLM Inference},pdfauthor={Yichen Long, Xiaoyu Hao, Hongyang Pan, Jianyi Cheng, Chen Bai, Keren Zhu},pdfsubject={},pdfkeywords={}}

\title{Upright DRAM on Logic: Edge-Connected Memory Cubes for Bandwidth-Bound LLM Inference}
\author{\IEEEauthorblockN{Yichen Long\textsuperscript{1}, Xiaoyu Hao\textsuperscript{1}, Hongyang Pan\textsuperscript{2}, Jianyi Cheng\textsuperscript{3}, Chen Bai\textsuperscript{4,*} and Keren Zhu\textsuperscript{1,*}}
\IEEEauthorblockA{\footnotesize\textsuperscript{1}College of Integrated Circuits and Micro-Nano Electronics and \textsuperscript{4}Institute of Systems for Advanced Computing, Fudan University, Shanghai, China\\
\textsuperscript{2}Shanghai Trace Intelligence Co., Ltd., Shanghai, China\quad
\textsuperscript{3}School of Informatics, University of Edinburgh, Edinburgh, UK\\
\{longyc26, xyhao26\}@m.fudan.edu.cn, panhongyang@traceic.com, jianyi.cheng@ed.ac.uk, \{cbai, krzhu\}@fudan.edu.cn\quad
\textsuperscript{*}Co-corresponding authors}}

\begin{document}
\bstctlcite{v3:BSTcontrol}
\maketitle

\begin{abstract}
Token generation in large language models is often limited by memory bandwidth, because every step fetches model weights and the key-value history. High-bandwidth memory (HBM) stacks dynamic random-access memory (DRAM) dies to meet this demand, yet HBM beside a single-reticle logic die is bounded by the die perimeter, and HBM above the logic faces coupled cooling and power-feed constraints. Upright packages such as Z-Angle Memory and massive orthogonal stacking assembly (MOSAIC) turn DRAM dies on edge to gain area-wide links and cooling gaps. However, one shared edge then carries power, commands and data, so heat and voltage drop govern bandwidth, and the many narrow channels may congest the on-chip mesh. The proposed processor therefore places liquid-cooled cubes of edge-bonded DRAM dies on a reticle-scale logic die and maps the data of each core to the channels above it. We size this design with a design space exploration that joins cooling, power distribution, wiring, assembly and inference latency. With demonstrated bump pitches, the coarse edge contacts hold the processor at 2.6\,TB/s against 8.9\,TB/s for HBM beside the logic. With projected pitches, it reaches 45.2\,TB/s, capped by supply voltage drop, against 14.9\,TB/s. For three models of 8 to 70 billion parameters targeting 12\,ms per token, it delivers 2.1--2.4$\times$ the decode throughput of HBM beside the logic with host step time calibrated to H100 runs (2.0--3.4$\times$ over 0--4\,ms), at similar first-token latency and equal compute.
\end{abstract}

\section{Introduction}
\label{sec:intro}

\chen{Large language model (LLM) inference is limited by memory bandwidth.}
\chen{Each decode step reads the weights once and the key--value (KV) cache of every sequence, so the time per output token (TPOT) scales inversely with memory bandwidth~\cite{zhang2024llmcompass}.}

\begin{figure}[!tb]
\centering
\includegraphics[width=\columnwidth]{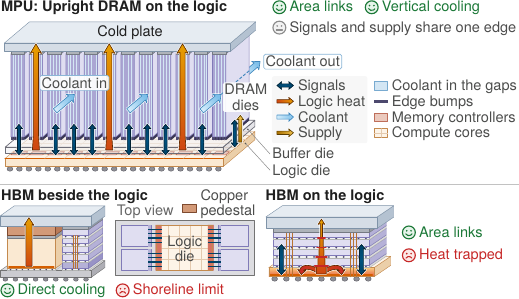}
\caption{Top: twelve cubes of upright DRAM dies on the buffer and logic dies; bottom left: HBM beside the logic die, where all memory links cross the shoreline, the edge of the logic die; bottom right: HBM on the logic die.}
\label{fig:mpu}
\end{figure}

\chen{High-bandwidth memory (HBM) stacks are built from DRAM dies interconnected by through-silicon vias (TSVs), and the stack sits on a silicon interposer beside the logic die (Fig.~\ref{fig:mpu}, bottom left).}
\chen{A reticle field, however, is only 26 by 33\,mm~\cite{xu2024opc}, so a single-reticle logic die offers at most 33\,mm of shoreline, \textit{i.e.},~the edge available for memory links.
Because that shoreline bounds the number of links, memory beside the die gains bandwidth mainly through higher link rates.}

\chen{Stacking memory and logic frees the shoreline but comes at a cost.
Depending on which sits on top, either the logic's heat or its power and signals must cross the memory.
For example, with HBM stacked on the logic die, the links run vertically, so bandwidth scales with the logic area rather than the shoreline.
The logic's heat, however, must then cross every DRAM die (Fig.~\ref{fig:mpu}, bottom right).
Prior studies have co-designed cooling and power delivery for 3D-stacked accelerators~\cite{sharda2025stacked,he2025tasa,lee2026coupled}, analyzed the cost of 3D-stacked LLM accelerators~\cite{bai2025accelstack} and proposed monolithic 3D DRAM~\cite{pan2025stratum}.
These studies exploit 3D-stacked DRAM through careful architecture and process design, but they do not remove the need for either the logic's heat or its power and signals to cross the entire memory die.
}

\chen{\textit{Upright memory} offers a different trade-off.
Its dies stand upright on the logic die, side by side, leaving gaps between them.
The heat of the logic can then escape upward through those gaps.
Each die is bonded to the logic die along its bottom edge, and the bonds form the links (Fig.~\ref{fig:mpu}, top).
Moreover, an upright die occupies little of the logic area, and its capacity grows with the height of the die.
Several projects are now building this architecture, including micro-bump massive orthogonal stacking assembly (MOSAIC)~\cite{huang2026first}, V-die~\cite{yang2026vertical}, contactless MOSAIC~\cite{mitarai2026orthogonal,kawano2025massive} and Z Angle Memory~\cite{morein2026challenges}.
To evaluate this memory in a system, a recent thermal study placed upright DRAM on a GPU and estimated the training speed assuming a bandwidth linear in the number of active dies~\cite{chen2026beyond}.
Focusing on the thermal envelope, it left edge links, supply voltage drop, coolant between dies and die assembly outside its scope.
These open issues are coupled, and in this work, we fill the gap by addressing them together.
To our knowledge, no prior work has co-designed a processor and its upright memory under these coupled limits.
}

\begin{figure*}[t]
\centering
\includegraphics[width=\textwidth]{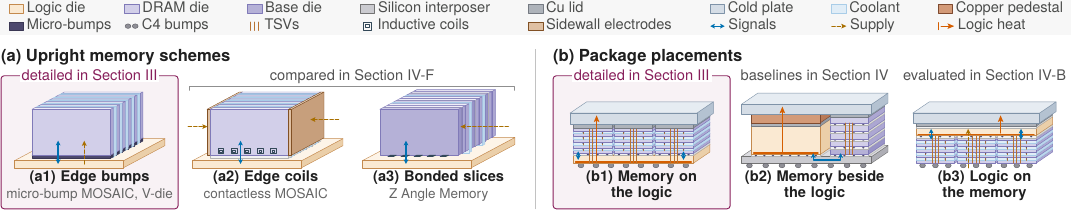}
\caption{Memory interfaces and package placements: (a)~three upright memory schemes, (a1)~edge bumps~\cite{huang2026first,yang2026vertical}, (a2)~edge coils~\cite{mitarai2026orthogonal,kawano2025massive} and (a3)~bonded slices~\cite{morein2026challenges,abedin2026induction}, which differ in how a die edge carries signals and supply, and (b)~three placements of memory with a logic die, drawn with HBM stacks.}
\label{fig:schemes}
\end{figure*}

\chen{We propose \SysName{}, a processor that places cubes of upright DRAM dies on a 22 by 30\,mm logic die.
The DRAM dies connect to the logic die through edge bumps, as in micro-bump MOSAIC~\cite{huang2026first}.
Each DRAM die carries eight channels through its bottom edge to controllers below (Fig.~\ref{fig:mpu}, top).
Because uniform address interleaving over 864 channels would overload the 16.9\,TB/s mesh bisection in the logic die, \SysName{} uses channel-local mapping.
Each core in the logic die then reads mainly from the dies above it.
Yet power supply and data signals share one edge, so the bandwidth limits itself.
With projected technology, we find that links that could carry 64.3\,TB/s deliver 45.2\,TB/s.
Co-designing \SysName{} therefore requires a design space exploration that couples thermal, power delivery, interconnect and assembly limits with LLM inference.
We make the following contributions.}

\begin{itemize}
\item We describe \SysName{}'s microarchitecture, package and channel-local mapping, which sends below 0.2\,TB/s across the die midline; uniform interleaving would send 22.6\,TB/s and cap \SysName{}'s lead at 1.7--2.1$\times$ (Section~\ref{sec:design}).
\item We present a design space exploration coupling heat, power delivery, links and assembly to find the bandwidth and binding limit of upright memory and HBM, on and beside the logic, under shared limits; on 21 unbuffered finite-element checks, its fast heat model is 56--12,750$\times$ faster and errs by at most 8.3 and 1.7\,K in the logic and DRAM, and its refined model by at most 1.4 and 1.1\,K, or 2.3\,K with the buffer (Section~\ref{sec:dse}).
\item We compare four configurations and three upright schemes: current coarse edge bumps hold \SysName{} to 2.6 against 8.9\,TB/s for HBM beside the logic, whereas with projected technology it delivers 45.2 against 14.9\,TB/s and serves 2.1--2.4$\times$ the decode throughput at 12\,ms TPOT with fitted host CPU time (2.0--3.4$\times$ for 0--4\,ms), with a similar time to first token at equal compute.
\end{itemize}

\section{Preliminaries}
\label{sec:prelim}

\subsection{Upright memory}
\label{sec:upright}

Upright memory is a new memory paradigm. Each DRAM die is rotated by 90$^\circ$ with one of its edges facing the substrate. The die's signal interface is placed along that edge, giving each die an independent datapath to the substrate. Since every die can now transfer data in parallel, adding dies increases both capacity and aggregate bandwidth. Moreover, because the datapath does not need to pass through adjacent dies, tight die-to-die bonding is no longer needed. This reduces the need for TSVs and allows optional gaps between adjacent dies for direct coolant contact (Fig.~\ref{fig:schemes}a). 
\color{black}

\color{black}

Reported upright designs fall into three schemes based on signal and power delivery. \textit{Edge bumps} (Fig.~\ref{fig:schemes}, a1) route both via micro-bumps along the bottom edge~\cite{huang2026first,yang2026vertical}. \textit{Edge coils} (a2) inductively couple signals without contact while supplying power through the sidewalls~\cite{mitarai2026orthogonal,kawano2025massive}. \textit{Bonded slices} (a3) face-bond DRAM dies to a logic base die, forming an upright slice whose base die drives edge coils~\cite{morein2026challenges,abedin2026induction}.

Edge bumps maximize edge link density while preserving inter-die coolant gaps absent in bonded slices: three rows at a projected 20\,\textmu m pitch pack 150 bumps/mm, versus roughly three coils/mm at a 320\,\textmu m pitch~\cite{mitarai2026analysis}. However, sharing the bottom edge for power increases supply voltage drop with die height, and each die requires sub-pitch alignment to the bump grid. Section~\ref{sec:design} details \SysName{}, and Section~\ref{sec:ablation} compares the three schemes.

\subsection{Stacked memory and bonding}
\label{sec:stacking}

HBM stacks dies face to back on TSVs and a base die, 12 in HBM3E and 16 in HBM4~\cite{samsung24_hbm3e_12h,jedec2025jesd270}, joined by micro-bumps or hybrid bonds (Fig.~\ref{fig:schemes}, b2). Micro-bumps at a 55\,\textmu m pitch~\cite{yang2026vertical} need underfill, whereas hybrid bonds join copper and oxide~\cite{workman2023hybrid,workman2026reliability}. Heat crossing the stack passes one bond layer per die, each adding thermal resistance~\cite{kim_ectc23,han_eptc21}.

Memory and logic stack in either order (Fig.~\ref{fig:schemes}, b1 and~b3), as products already stack logic on logic~\cite{mathur2020thermal}. We evaluate logic on memory only with HBM, as upright dies lack vertical TSVs for logic supply and already give heat a vertical path.

\subsection{Large language model inference}
\label{sec:llm}

LLM inference runs in two phases, a compute-bound prefill and a bandwidth-bound decode~\cite{patel2024splitwise}, and each phase sets one latency. Prefill processes prompt tokens together and reuses each weight across them, whereas decode generates one token per sequence per step and reads all weights, shared by the batch, and each sequence's KV cache, which gives at most $2N$ operations per 8-bit weight byte at batch $N$ but a fixed number per KV cache byte. The time to the first token (TTFT) covers the prefill and the TPOT each later step. A larger batch $N$ raises the throughput ($N/\mathrm{TPOT}$), but its KV cache reads lengthen each step, so a faster memory raises it at a target rate per user ($1/\mathrm{TPOT}$) without raising the TTFT.

\section{\SysName{} Architecture and Design Space Exploration}
\label{sec:design}

\subsection{Overview}
\label{sec:overview}

\begin{figure}[!t]
\centering
\includegraphics[width=\columnwidth]{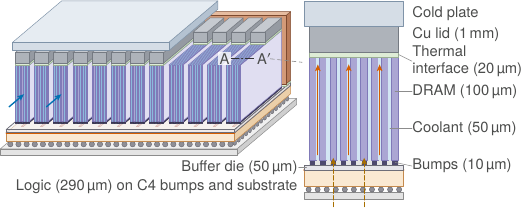}
\caption{Package of \SysName{} with projected technology: oblique view of the twelve cubes on the logic die, cold plate and copper pedestals partly cut away (left), and cross section A--A$'$ of one cube (right); layer heights are schematic.}
\label{fig:package}
\end{figure}

\begin{figure*}[t]
\centering
\includegraphics[width=\textwidth]{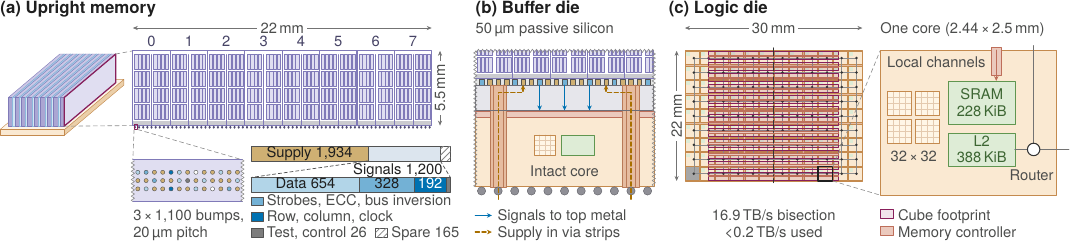}
\caption{Microarchitecture of \SysName{} with projected technology: (a)~an upright DRAM die, its bottom-edge micro-bumps colored by class share, not position, and bump sites per HBM3 signal class~\cite{jedec2025jesd238}; (b)~a buffer-die section with vertical signals; (c)~the logic die with cube footprints, 107 cores and one spare.}
\label{fig:micro}
\end{figure*}

\SysName{} is designed to exploit the strengths of upright memory while managing its thermal challenges (Fig.~\ref{fig:package}).
On top, upright memory forms an array. In the design of Section~\ref{sec:dse}, twelve \textit{cubes} stand on the logic die. Each cube is 22\,mm long, 5.5\,mm tall and about 1.35\,mm thick. It contains nine DRAM dies, and we assume that sealant along their top and bottom edges holds them together. A 50\,\textmu m gap separates adjacent dies, so coolant flows past all dies, while an assumed dielectric coating seals the bumps. The cubes stand in a row along the 22\,mm side of the logic die, with their long sides running parallel to the logic die's 30\,mm side. A 0.5\,mm clearance separates neighboring cubes. Each cube carries a thermally conductive lid, and one flat cold plate covers all the lids.
At the bottom lies the 22 by 30\,mm tiled logic die (Fig.~\ref{fig:micro}c).

Between the memory array and the logic die lies a 50\,\textmu m buffer die (Fig.~\ref{fig:micro}b). It takes the 20\,\textmu m-pitch micro-bumps of the upright dies on its top face and connects to the logic die through hybrid bonds on its bottom face~\cite{workman2023hybrid,smith2024realizing}. The buffer die routes signals to the receivers and controllers under each cube. It also gathers the supply into via strips between the cores, so no supply via crosses a core.

\subsection{Microarchitecture of the upright memory}
\label{sec:memarch}

Fig.~\ref{fig:micro}a depicts the microarchitecture of the upright die. In the design selected in Section~\ref{sec:dse}, each die is partitioned into eight independent channels. Each channel has dedicated row, column and clock signals and a controller on the logic die below.

Along the bottom edge, three rows of 20\,\textmu m-pitch micro-bumps provide 3,300 sites. Of these, 1,934 are allocated to supply and ground, 192 to row, column and clock signals and 328 to strobe, ECC and bus-inversion signals. Following HBM3~\cite{jedec2025jesd238}, the latter two groups require 24 sites per channel and one site per two data signals, respectively. Another 26 sites support test and control, 165 serve as spares, and 654 transfer data.
Each channel stores roughly 3.6\,Gibit and has about 82 data signals, so each die's data interface delivers 654\,GB/s at 8\,Gb/s per signal, and a nine-die cube integrates 72 channels and 32.8\,GiB. Below 85\,$^\circ$C, refresh blocks all banks of a channel simultaneously for 350\,ns every 3.9\,\textmu s~\cite{jedec2021jesd235d}, \textit{i.e.},~9\% of the time, and twice as often above 85\,$^\circ$C. Due to refresh, a die's arrays supply 678 of their 745\,GB/s, and the links of twelve cubes could sustain 64.3\,TB/s. However, supply voltage drop caps the delivered bandwidth (Section~\ref{sec:bw}).

\begin{table}[t]
\centering
\caption{Design variables and limits of the exploration, for upright memory and for HBM, on and beside the logic die.}
\label{tab:space}
\footnotesize
\setlength{\tabcolsep}{3pt}
\begin{tabularx}{\columnwidth}{@{}>{\raggedright\arraybackslash}p{0.27\columnwidth}>{\raggedright\arraybackslash}X>{\raggedright\arraybackslash}X@{}}
\toprule
 & Upright memory & HBM \\
\midrule
Dies & 1--16 cubes of 4--48, 8--22 by 5.5--11\,mm & 1--16 stacks of 4--16, 8--12.5\,mm wide \\
Die thickness & 50--300\,\textmu m & 50--300\,\textmu m \\
Gap between dies & 5 or 50\,\textmu m, optional coolant & bond layer, or a coolant layer as in~\cite{lee2026coupled} \\
Supply feed & bottom edge or sidewalls & TSVs \\
Channels, rate & 2, 4, 6 or 8 per die, 4 or 8\,Gb/s & product HBM3E and HBM4 stacks \\
\midrule
Limits & \multicolumn{2}{>{\raggedright\arraybackslash}p{0.68\columnwidth}}{DRAM and logic temperature, coolant flow and pressure, voltage drop and contact current, edge sites or shoreline, receiver area, edge alignment, assembly height, capacity}\\
\bottomrule
\end{tabularx}
\end{table}

\subsection{Logic die and channel-local mapping}
\label{sec:logicarch}

The $22 \times 30$\,mm logic die (Fig.~\ref{fig:micro}c) integrates a $9 \times 12$ tile grid comprising 107 cores and one spare. Each core features four matrix units of $32 \times 32$ 8-bit multiply-accumulate cells, 228\,KiB of static random-access memory (SRAM) and a 388\,KiB L2 cache slice. Area-scaling an H100~\cite{nvidia2026h100,andersch2022hopper} yields this L2 capacity and 1,605 8-bit TFLOPS at 1.83\,GHz within 90\,W/cm$^2$. Because upright memory channels are distributed across the die, most cores reside directly beneath the channels they access, enabling all cores to read in parallel. Receivers for the 129,600 edge signals occupy 15.6\% of the cube footprints, \textit{i.e.},~55.6\,mm$^2$ or 8.4\% of the die.

Interleaving data across these channels would induce heavy cross-die traffic, so \SysName{} adopts channel-local mapping instead. Following tensor parallelism~\cite{shoeybi2019megatron}, software partitions the output and down projections by input columns and the remaining weights by output rows, ensuring each core primarily fetches its own weights from overlying channels. Each half of the logic die stripes the KV cache of half the heads across its cores in 16-token blocks, confining partial attention merges within that half. Each decode step thus utilizes all channels up to the delivered bandwidth $B^\ast$, while the busiest core reads under 1\% more bytes than average. Per layer, only activations and partial sums traverse the mesh bisection. For the largest model at batch size 56, they total 3.7\,MB, versus 1.8\,GB of weights and KV cache read per layer. In contrast, uniform interleaving would route 22.6\,TB/s across this 16.9\,TB/s bisection. Under an analytical bound, this would cap \SysName{} at 33.7\,TB/s and its lead at 1.7--2.1$\times$.

\subsection{Multi-physics design space exploration}
\label{sec:dse}

Finding the best design requires enforcing shared physical constraints. As each DRAM die dissipates $P_\text{die}=P_\text{bg}+(E_\text{array}+E_\text{link})B_\text{die}$ at bandwidth $B_\text{die}$, heat and voltage drop grow with the total bandwidth $B$. Once either binds, adding bumps cannot raise the bandwidth, so the memory-die count, size and spacing, supply feed, cooling and links must be chosen jointly (Table~\ref{tab:space}). Because finite-element simulation is too slow for this large, coupled space, we pair fast screening models with selective thermal refinement. The models return normalized constraints ($g\le 0$ is feasible) and bandwidth caps $c_m$, from which Algorithm~\ref{alg:search} scores each design by its delivered bandwidth
\begin{equation}
B^\ast=\max\{B\le\min_m c_m\colon g_\text{heat}(B),g_\text{supply}(B),g_\text{fit}\le 0\},
\label{eq:bw}
\end{equation}
where $g_\text{fit}$ groups the $B$-independent constraints of geometry $g_\text{geo}$, assembly and capacity.

Four fast models evaluate each design in closed form or as a compact network. A thermal network along each upright die or HBM stack captures laminar gap heat transfer and heat removal by the gap flow and cold plate over an insulated substrate. The other three estimate the voltage drop of a TSV or edge feed, which for an edge feed grows with die height over width, count bump sites per edge or along the shoreline and estimate the probability that all edges align within a quarter of the bump pitch~\cite{mitarai2026analysis}.

In each scenario, the multi-objective genetic search NSGA-II~\cite{deb2002fast,blank2020pymoo} uses the fast models to maximize $B^\ast$ and capacity $M$ while minimizing energy per bit $E_\text{bit}$ and footprint $A$ (Algorithm~\ref{alg:search}). The refined heat model then re-evaluates the best design and the leading designs near thermal limits, subdividing only the grid of the thin logic layer and contact edges without fitting coefficients. Across 21 finite-element checks without the buffer die, the fast model errs in peak temperature by at most 8.3\,K in the logic (2.9\,K on average) and 1.7\,K in the DRAM while running 56--12,750$\times$ faster. The refined model cuts these maxima to 1.4 and 1.1\,K (Section~\ref{sec:thermal}).

\subsection{Inference performance model}
\label{sec:package}

\begin{algorithm}[t]
\caption{Bandwidth of one design and two-level search.}
\label{alg:search}
\footnotesize
\algrenewcommand\algorithmicindent{1em}
\begin{algorithmic}[1]
\Function{Bw}{$x,h$} \Comment{heat model $h$, $g=\max(g_\text{heat},g_\text{supply})$}
  \State \algorithmicif\ $g_\text{geo}(x)>0$ \algorithmicthen\ \Return infeasible
  \State $B_l,B_u\gets 0,\min_m c_m$;\; \algorithmicif\ $g(B_u)\le 0$ \algorithmicthen\ $B_l\gets B_u$
  \State \algorithmicelse\ \algorithmicif\ $g(0)>0$ \algorithmicthen\ \Return infeasible \Comment{$P_\text{die}=P_\text{bg}$}
  \While{$B_u-B_l>\epsilon$} \Comment{$P_\text{die}$ grows with $B$}
    \State $B\gets(B_l+B_u)/2$;\; \algorithmicif\ $g(B)\le 0$ \algorithmicthen\ $B_l\gets B$ \algorithmicelse\ $B_u\gets B$
  \EndWhile
  \State \Return $B_l$ if $g_\text{fit}\le 0$, else infeasible
\EndFunction
\State $\mathcal{F}\gets$ NSGA-II on $\textsc{Bw}(x,\text{fast})$: $\max\,(B^\ast,M)$, $\min\,(E_\text{bit},A)$
\State $\mathcal{X}=\{x_1,\dots,x_{50}\}\gets$ best of $\mathcal{F}$ by $B^\ast$, then $E_\text{bit}$
\State $\mathcal{R}\gets\{x_1\}\cup\{x\in\mathcal{X}:$ within 10\,K (logic), 3\,K (DRAM) of a limit$\}$
\State \algorithmicrepeat\ update $x\in\mathcal{R}$ by $\textsc{Bw}(x,\text{refined})$;\; $x^\ast\gets$ best feasible in $\mathcal{X}$
\State $\mathcal{R}\gets\{x^\ast\}$ \algorithmicuntil\ $x^\ast$ is refined;\; \Return $x^\ast$ \Comment{by $B^\ast$, then $E_\text{bit}$}
\end{algorithmic}
\end{algorithm}

An analytical simulator turns the delivered bandwidth $B^\ast$ of each design into prefill and decode step times. With compute overlapping reads and a barrier after each kernel, a step takes
\begin{equation}
t=t_\text{head}+t_o+\sum_{\ell\in\mathcal{L}}\sum_{k\in\mathcal{K}_\ell}\max\Big(\frac{O_k}{\eta_kF},\,\frac{S_k/\eta_s+G_k/\eta_g}{B^\ast}\Big),
\label{eq:step}
\end{equation}
where kernel $k$ of layer $\ell$ performs $O_k$ operations at efficiency $\eta_k$ of peak compute $F$, streams $S_k$ contiguous and gathers $G_k$ paged bytes, with $t_\text{head}$ the output-layer time and $t_o$ the host CPU time per step. Efficiencies are constants of the MemExplorer simulator~\cite{wu2026memexplorer}, except that we divide its $\eta_s=0.85$ and $\eta_g=0.45$ by 0.91, removing the 9\% refresh loss already in $B^\ast$. With bisection traffic below 0.2\,TB/s and core byte loads within 1\% (Section~\ref{sec:logicarch}), we use $B^\ast$ without weight caching (64\,MiB on chip, under 1\% of weights).

\color{black}

\section{Experiments}
\label{sec:eval}

\subsection{Setup}
\label{sec:setup}

\begin{figure}[!t]
\centering
\includegraphics[width=\columnwidth]{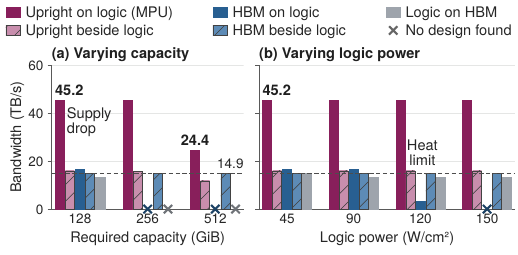}
\caption{Best projected bandwidth of five layouts against (a)~capacity at 90\,W/cm$^2$ and (b)~logic power at 128\,GiB; at 512\,GiB \SysName{} has 8\,mm tall dies.}
\label{fig:walls}
\end{figure}

\begin{figure*}[t]
\centering
\includegraphics[width=\textwidth]{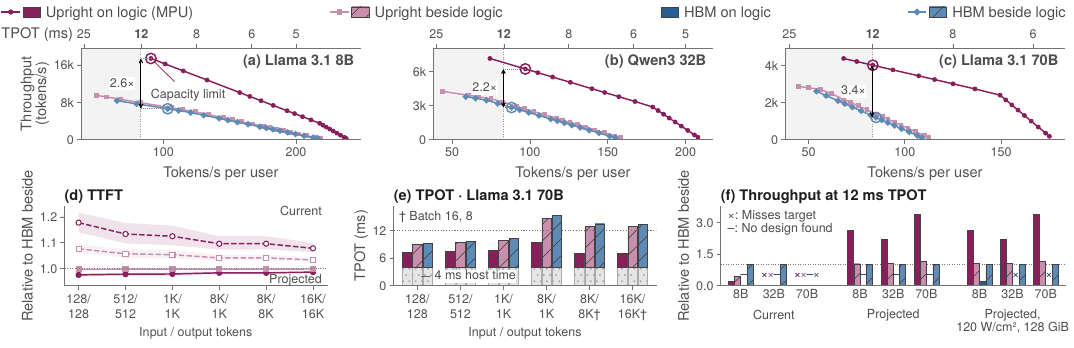}
\caption{Inference performance of the four configurations: (a)--(c)~throughput against the rate per user with projected technology at 90\,W/cm$^2$ and 256\,GiB, 8K/1K tokens; (d)~TTFT and (f)~throughput at the 12\,ms TPOT target relative to HBM beside the logic; (e)~TPOT of the largest model; 4\,ms host time.}
\label{fig:perf}
\end{figure*}
\begin{figure*}[t]
\centering
\begin{minipage}[b]{0.485\textwidth}\centering
\includegraphics[width=\linewidth]{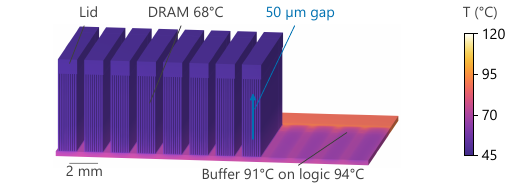}
\end{minipage}\hfill
\begin{minipage}[b]{0.485\textwidth}\centering
\includegraphics[width=\linewidth]{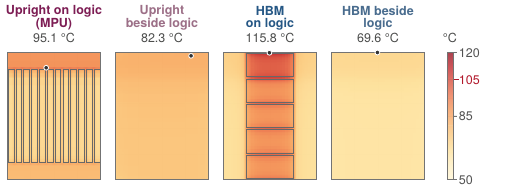}
\end{minipage}
\caption{Decode temperature (128/1K tokens, batch 32) of the 70 billion parameter model with projected technology at 120\,W/cm$^2$ and 128\,GiB, with logic power split into compute and SRAM. Left: \SysName{} by finite elements, cut midway through the dies, with four cubes, pedestals and interface hidden from view (45 to 120\,$^\circ$C). Right: the logic's top face by the refined model, with memory outlines, peak dots and HBM on the logic limited to 3.1\,TB/s (50 to 120\,$^\circ$C).}
\label{fig:thermal}
\end{figure*}
To isolate memory type and placement, we optimize \SysName{} (upright memory with edge bumps on logic) and three variants, moving memory beside logic, switching to HBM or both, under identical parameters and limits (Table~\ref{tab:best}) on a single-reticle logic die, at 8\,Gb/s/pin in projected technology. HBM beside logic (as in H100~\cite{andersch2022hopper}) is our main baseline; stitching dies scales on-logic cubes with area but shoreline HBM only with perimeter. A fifth layout, logic on HBM (Fig.~\ref{fig:walls}), fits at 128\,GiB but not 256\,GiB.

Three scenarios pair current technology (hardware-demonstrated links and stacks) or projected technology (adding specifications and roadmaps) with logic power density and capacity: current and projected at 90\,W/cm$^2$ with 256\,GiB, and projected at 120\,W/cm$^2$ with 128\,GiB. All use one flat cold plate (100{,}000\,W/m$^2$K, 45\,$^\circ$C water) on copper pedestals.

In the simulator of Section~\ref{sec:package}, we run 8-bit Llama~3.1 (8B, 70B)~\cite{grattafiori2024llama} and Qwen3 (32B)~\cite{yang2025qwen3}. Fig.~\ref{fig:perf}a--c sweeps batch sizes 1--512 at 8K/1K input/output tokens, comparing throughput at an interactive 12\,ms TPOT (83 tokens/s/user); other runs use batch 32 (or the largest fitting power of two) across 128/128--16K/16K, while thermal and power runs use prefill- and decode-dominated 8K/128 and 128/1K tokens. Across 11 H100 runs of the 8B model at batch 1, the simulator with its 4\,ms default host time is within $-17$ to $+13$\% on TTFT. On TPOT, omitting host time underestimates all runs ($-60$ to $-12$\%), whereas 4\,ms overestimates 10 of 11 ($-25$ to $+67$\%) because 10 measured steps last only 4.4--6.5\,ms; a fitted 1.95\,ms gives zero mean error ($-43$ to $+27$\%). We use 1.95\,ms unless stated (4\,ms in Figs.~\ref{fig:perf}--\ref{fig:ablation}) and compare ratios.

\color{black}
\subsection{Bandwidth across capacity and power}
\label{sec:bw}

\begin{table}[!t]
\centering
\caption{Best designs of the four configurations in three scenarios.}
\label{tab:best}
\scriptsize
\setlength{\tabcolsep}{2.75pt}
\begin{tabular}{@{}l l r r l r r@{}}
\toprule
& & Bandwidth & Capacity & & \multicolumn{2}{c}{Peak ($^\circ$C)}\\
\cmidrule(l){6-7}
Scenario & Configuration & (TB/s) & (GiB) & Limit & DRAM & Logic\\
\midrule
\multirow{5}{*}{\makecell[l]{Current\\90\,W/cm$^2$\\256\,GiB}} & \makecell[tl]{\textcolor{plUpOn}{$\blacksquare$}\,Upright on logic\\\hphantom{$\blacksquare$}\,(MPU)} & 2.6 & 326 & links & 62 & 74\\
 & \textcolor{plUpBeside}{$\blacksquare$}\,Upright beside & 4.3 & 552 & voltage drop & 70 & 79\\
 & \textcolor{plHbmOn}{$\blacksquare$}\,HBM on logic & \multicolumn{5}{c}{---}\\
 & \textcolor{plHbmBeside}{$\blacksquare$}\,HBM beside & 8.9 & 288 & links & 64 & 62\\
\midrule
\multirow{5}{*}{\makecell[l]{Projected\\90\,W/cm$^2$\\256\,GiB}} & \makecell[tl]{\textcolor{plUpOn}{$\blacksquare$}\,Upright on logic\\\hphantom{$\blacksquare$}\,(MPU)} & 45.2 & 393 & voltage drop & 67 & 86\\
 & \textcolor{plUpBeside}{$\blacksquare$}\,Upright beside & 15.8 & 339 & arrays & 58 & 81\\
 & \textcolor{plHbmOn}{$\blacksquare$}\,HBM on logic & \multicolumn{5}{c}{---}\\
 & \textcolor{plHbmBeside}{$\blacksquare$}\,HBM beside & 14.9 & 256 & arrays & 63 & 62\\
\midrule
\multirow{5}{*}{\makecell[l]{Projected\\120\,W/cm$^2$\\128\,GiB}} & \makecell[tl]{\textcolor{plUpOn}{$\blacksquare$}\,Upright on logic\\\hphantom{$\blacksquare$}\,(MPU)} & 45.2 & 393 & voltage drop & 70 & 95\\
 & \textcolor{plUpBeside}{$\blacksquare$}\,Upright beside & 15.9 & 175 & links & 59 & 83\\
 & \textcolor{plHbmOn}{$\blacksquare$}\,HBM on logic & 3.1 & 146 & temperature & 50 & 105\\
 & \textcolor{plHbmBeside}{$\blacksquare$}\,HBM beside & 14.9 & 203 & arrays & 51 & 71\\
\bottomrule
\end{tabular}

\vspace{2pt}
\parbox{\columnwidth}{\scriptsize Temperatures assume evenly spread logic power; a dash marks no feasible design found.}
\end{table}
With current technology, the coarse edge-bump pitch caps \SysName{} at 2.6\,TB/s, below upright memory (4.3\,TB/s) and HBM (8.9\,TB/s) beside the logic (Table~\ref{tab:best}). With projected technology, \SysName{} reaches 45.2\,TB/s, and its bottleneck shifts to supply voltage drop. Its 1.1\,V supply draws 1{,}340\,A in total through the 290\,\textmu m logic die, the buffer die and die copper of 0.015\,$\Omega$ per square, so the supply and ground drop reaches the 5\% limit at 13.7\,W and 418\,GB/s per die (4.24\% in the die copper, 0.58\% in the logic die, 0.11\% in the buffer die and 0.07\% in the contacts). Beside the logic, upright memory (15.8\,TB/s) and eight HBM4 stacks (14.9\,TB/s after refresh, with 2{,}048 pins at 8\,Gb/s~\cite{jedec2025jesd270}) are both limited by their DRAM arrays. Without memory beside its logic die, \SysName{} achieves 0.74$\times$, 8.6$\times$ and 5.8$\times$ the bandwidth per unit footprint of HBM beside the logic in the three scenarios of Table~\ref{tab:best}.

Across 48 settings that sweep power density (45--150\,W/cm$^2$), capacity (up to 512\,GiB) and cold plates (Fig.~\ref{fig:walls}), HBM leads in all 24 current-technology cases, whereas \SysName{} leads in 18 of the 24 projected cases. The exceptions use the 40{,}000\,W/m$^2$K cold plate at 512\,GiB or 150\,W/cm$^2$. At 512\,GiB, \SysName{} needs 8\,mm tall dies, whose longer vertical supply path caps it at 24.4\,TB/s (Fig.~\ref{fig:walls}a).

Stacking HBM and logic hits thermal or array limits. Heat restricts HBM4 on the logic to half the capacity (below 256\,GiB, Fig.~\ref{fig:walls}a) and caps it at 3.1\,TB/s at 128\,GiB and 120\,W/cm$^2$, whereas \SysName{} maintains 45.2\,TB/s (Fig.~\ref{fig:walls}b). Logic on HBM avoids this thermal limit, but its DRAM arrays cap it at 13.2\,TB/s (146\,GiB), below HBM beside the logic.

\subsection{Inference performance}
\label{sec:perf}

With projected technology, \SysName{} removes the edge-link bottleneck and delivers the highest throughput across per-user rates (Fig.~\ref{fig:perf}a--c). At the 12\,ms TPOT target, it achieves 2.3$\times$, 2.1$\times$ and 2.4$\times$ the throughput of HBM beside logic on the 8B, 32B and 70B models (2.1--2.4$\times$ without host time, and 2.6$\times$, 2.2$\times$ and 3.4$\times$ at 4\,ms), as weight reads and host overhead consume more of each HBM step (Fig.~\ref{fig:perf}e) and restrict concurrency (24 versus 56 sequences on 70B). On 8B, capacity rather than TPOT caps \SysName{} at 192 sequences, the largest sampled batch below the 234 that fit with KV cache, headroom and workspace. Capping its 393\,GiB at HBM's 256\,GiB (Table~\ref{tab:best}) still yields 2.23$\times$ at 4\,ms. Both fixed-batch TPOT speedup (1.06--1.89$\times$) and throughput gain stay below the 3.0$\times$ bandwidth gain. Host time and compute bound fixed-batch steps, and batches above 16 at the 12\,ms target turn \SysName{}'s weight products compute-bound. Such kernels take 30\%, 48\% and 61\% of its kernel time, versus 12\%, 0\% and 0\% on HBM beside logic.

With current technology, coarse edge links bottleneck \SysName{} (Fig.~\ref{fig:perf}f), so it meets the 12\,ms TPOT only on 8B, at 0.21$\times$ HBM beside logic. At 120\,W/cm$^2$ with projected technology, \SysName{} retains its lead, whereas thermally limited HBM on logic meets the target only on 8B.

Because shared logic compute bounds prefill, TTFT varies little (Fig.~\ref{fig:perf}d). With projected technology, both upright configurations stay within 3.2\% of HBM beside logic, while lower bandwidth adds 6--20\% for HBM on logic and 6--22\% for \SysName{} with current technology.
\color{black}

\subsection{Thermal analysis}
\label{sec:thermal}

\SysName{} keeps its logic cooler than HBM on the logic, as heat rises beside its dies, not through them (Fig.~\ref{fig:thermal}, left). With projected technology at 120\,W/cm$^2$, the logic's top face peaks in decode (2--7\,K above prefill) at 95\,$^\circ$C in \SysName{} (10\,K below the 105\,$^\circ$C limit, or 21\,K below at 90\,W/cm$^2$) and at 116\,$^\circ$C under HBM, against 70 and 82\,$^\circ$C beside HBM and upright memory (Fig.~\ref{fig:thermal}, right). Finite elements put \SysName{}'s logic 2.3\,K below the refined model, with its buffer die 3\,K below the logic and DRAM at 68\,$^\circ$C.
The search checks every limit at full bandwidth, where \SysName{} peaks at 70 and 95\,$^\circ$C in DRAM and logic (Table~\ref{tab:best}), so memory idle time relaxes none; there its coolant carries 85\% of the 2.4\,kW at 3.2\,L/min, warming 9.6\,K at a 10\,kPa pressure drop up a 5.5\,mm pass, or 160\,kPa along the 22\,mm dies as drawn, at equal temperatures. The search holds HBM on the logic at 105\,$^\circ$C with uniform power, but decode puts 25\% of logic power on SRAM (20\% of the die), which would throttle HBM further. A cold plate stepped to meet each lid~\cite{ahuja2014integrated,boudreaux2004adjustable} would cool \SysName{}'s logic by 7--10\,K at equal bandwidth and raise HBM on the logic from 3.1 to 7.3\,TB/s.

\subsection{Power and energy}
\label{sec:power}

\begin{figure}[!t]
\centering
\includegraphics[width=\columnwidth]{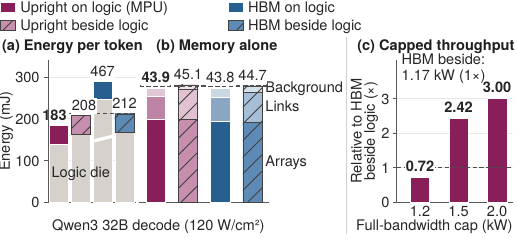}
\caption{(a)~Decode energy per token, (b)~memory part, Qwen3 32B, 128/1K tokens, batch 32, 120\,W/cm$^2$; (c)~throughput under full-bandwidth power caps, Llama 3.1 70B, 8K/1K tokens, 12\,ms TPOT, 90\,W/cm$^2$; projected technology.}
\label{fig:power}
\end{figure}

The logic die draws most of the power, so a faster memory can save decode energy despite drawing more power (Fig.~\ref{fig:power}a). We fix logic power at 792\,W, an upper bound as decode idles much of its compute, and memory draws at most 6\% in 8K/128 prefill and 33\% in decode. Per output token, \SysName{}'s memory spends 43.9 against 44.7\,mJ for HBM beside the logic (Fig.~\ref{fig:power}b), as both read the same bytes and its links spend 1.0 instead of 1.3\,pJ/b without the lateral interposer wires beside the logic. \SysName{} thus spends 13\% less energy per token (18\% at 1.95\,ms), mostly by running the logic die for less time, while HBM on the logic spends 2.2$\times$ as much as HBM beside it. During decode, \SysName{} averages 1.04 of its 2.45\,kW full-bandwidth power, as the 4\,ms host time idles its memory for 71\% of each step. At 90\,W/cm$^2$ it draws 2.25\,kW at 45.2\,TB/s and HBM beside the logic 1.17\,kW at 14.9\,TB/s, so a common cap of 1.2, 1.5 or 2.0\,kW limits only \SysName{}, to 13.6, 24.4 or 37.3\,TB/s, and gives the largest model 0.72$\times$, 2.4$\times$ and 3.0$\times$ the throughput with 4\,ms host time (Fig.~\ref{fig:power}c) and 0.83$\times$, 1.8$\times$ and 2.2$\times$ with 1.95\,ms. Against two baseline packages (2.35\,kW) in ideal tensor parallelism, with twice the compute and no communication cost, it serves 1.2$\times$, 1.0$\times$ and 0.9$\times$ as much for the three models.

\subsection{Comparison of upright memory schemes}
\label{sec:ablation}

\begin{figure}[t]
\centering
\includegraphics[width=\columnwidth]{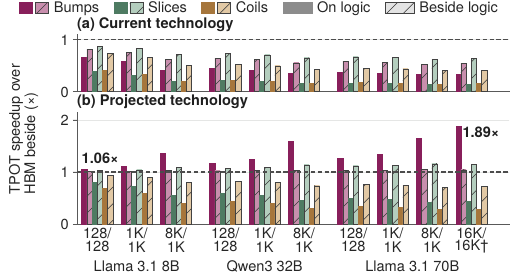}
\caption{TPOT speedup of three upright schemes (on and beside logic) over HBM beside logic (dashed) with (a)~current and (b)~projected technology at 90\,W/cm$^2$ and 256\,GiB, across 10 of 17 model--length pairs at batch 32 (batch 8 at $\dagger$).}
\label{fig:ablation}
\end{figure}

We compare three upright schemes, edge bumps, bonded slices and edge coils, on and beside the logic. Because they hit different bottlenecks, slices lead beside logic and edge bumps lead on it (Fig.~\ref{fig:ablation}). Each slice drives inductive coils from a logic base die~\cite{abedin2026induction}, modeled with 16 data coils of 26~\cite{ditzel2014thruchip} at 8.5 and 12.8\,Gb/s (current and projected) from stacked-logic measurements~\cite{shiba2023manchester,shiba2023encodingless}. Beside logic, coil-limited slices reach 5.2 and 18.3\,TB/s, ahead of edge bumps at 4.3 and 15.8\,TB/s (limited by voltage drop and arrays). On logic, edge bumps (\SysName{}) lead at 45.2\,TB/s with projected technology versus 5.2\,TB/s for slices, as each slice has a fixed coil bandwidth and only 256 slices fit on logic against 896 beside it. Edge coils at a 320\,\textmu m pitch~\cite{mitarai2026analysis} and 4 or 8\,Gb/s~\cite{mitarai2026orthogonal} reach 1.1 and 3.4\,TB/s on logic. Over all 17 model--length pairs, \SysName{}'s TPOT speedup over HBM beside logic is 1.06--1.89$\times$ (projected) and 0.33--0.66$\times$ (current). For the 70B model at 12\,ms TPOT (projected), slices beside logic serve 1.7$\times$ HBM's throughput with 4\,ms host time (1.3$\times$ without host time), whereas on-logic slices and edge coils miss the target.

\begin{table}[!t]
\centering
\caption{Sensitivity of \SysName{} to one change at a time.}
\label{tab:sens}
\scriptsize
\setlength{\tabcolsep}{0pt}
\begin{tabular*}{\columnwidth}{@{\extracolsep{\fill}}l c c c r@{}}
\toprule
 & \SysName{} & HBM beside & \SysName{} & \multicolumn{1}{c}{Throughput}\\
Change & (TB/s) & (TB/s) & limit & \multicolumn{1}{c}{\SysName{}/HBM}\\
\midrule
None (Table~\ref{tab:best}) & 45.2 & 14.9 & voltage drop & 2.1--2.4$\times$\\
Supply copper, 0.02\,$\Omega$ per square & 35.2 & 14.9 & voltage drop & 1.7--2.2$\times$\\
Supply copper, 0.03\,$\Omega$ per square & 24.3 & 14.9 & voltage drop & 1.4--1.8$\times$\\
Supply copper, 0.05\,$\Omega$ per square & 14.8 & 14.9 & voltage drop & 1.0$\times$\\
No logic vias or buffer die & 52.6 & 14.9 & voltage drop & 2.2--2.5$\times$\\
HBM at 10\,Gb/s, faster arrays & 45.2 & 18.6 & voltage drop & 1.7--1.9$\times$\\
HBM at 12\,Gb/s, faster arrays & 45.2 & 22.4 & voltage drop & 1.4--1.7$\times$\\
Logic die 100\,\textmu m thick & 49.0 & 14.9 & voltage drop & 2.1--2.5$\times$\\
Dielectric coolant & 45.1 & 14.9 & voltage drop & 2.1--2.4$\times$\\
\quad at 120\,W/cm$^2$, 128\,GiB & 26.9 & 14.9 & logic temperature & 1.5--1.9$\times$\\
\bottomrule
\end{tabular*}

\vspace{2pt}
\parbox{\columnwidth}{\scriptsize Projected technology, 90\,W/cm$^2$, 256\,GiB unless noted; throughput at 12\,ms TPOT.}
\end{table}

\textit{Sensitivity.} \SysName{}'s bottleneck shifts from edge links to voltage drop as bumps shrink, and to logic temperature with dielectric coolant at 120\,W/cm$^2$ (Table~\ref{tab:sens}). Thinner supply copper erodes the lead most, to parity at 0.05\,$\Omega$ per square. Our assumed maximum of 0.03\,$\Omega$ per square equals reserving half the 5\% drop for transients.

\textit{Limitations.} We omit transient ($L\,di/dt$) drop, buffer-die decoupling capacitors, the cost of the added die copper, edge and buffer-die routing, coil crosstalk, bump fatigue, die bending, coolant sealing, coating thermal resistance, gap clogging, known-good-cube tests and rework. Carrier silicon, used for smaller steps in MI300~\cite{smith2024realizing}, would cap our 90\,W/cm$^2$ design at 30.9\,TB/s and overheat the 120\,W/cm$^2$ one. Deducting 8.4\% of logic compute for receivers (15\% with controllers and TSVs) still leaves 2.0--2.3$\times$ (1.8--2.2$\times$) HBM's throughput at up to 1.08$\times$ (1.16$\times$) its TTFT. Finally, we evaluate only dense 8-bit models without latent attention.
\color{black}

\section{Conclusion}
\label{sec:conclusion}

\SysName{} places upright DRAM dies above the cores that read them. With projected technology, it delivers 3.0$\times$ the bandwidth and 2.1--2.4$\times$ the decode throughput of HBM beside the logic.
\section*{Content Generated by AI}
Claude Opus 5.5 generated the simulation and plotting code and initial drafts of all sections. The authors verified all results, revised the manuscript and take full responsibility for the content.

\clearpage
\bibliographystyle{IEEEtran}
\bibliography{refs}
\end{document}